\documentclass[12pt,final]{amsart}

\usepackage{graphicx} 
\usepackage[margin=0.75in]{geometry}
\usepackage{amsmath}

\usepackage{verbatim}
\usepackage{amssymb}
\usepackage{amscd}

\theoremstyle{definition}

\theoremstyle{remark}

\author{Tatyana Barron}
\address{Department of Mathematics, University of Western Ontario, London Ontario N6A 5B7, Canada}
\email{tatyana.barron@uwo.ca}

\title[Dimensionality increase]{Dimensionality increase for error correction in the interaction between information space and the physical world}

\begin{document}

\maketitle

\begin{abstract}
The evolution of human intelligence led to the huge amount of data in the information space. Accessing and processing this data helps in finding solutions to applied problems based on finite-dimensional models. We argue, that formally, such a mathematical model can be embedded into a higher-dimensional model inside of which a desired solution will exist. 
In our model, the physical world and the information space are submanifolds of infinite-dimensional Hilbert spaces, and the processes, including information transmission, are maps between the submanifolds of the physical world or of the information space. We discuss how our perspective fits in the context of existing literature.  
Our theorem states that a submanifold in the parameter space of the physical world 
can be deformed to a target submanifold outside that space, with an appropriate count of the deformation parameters. We interpret this assertion as an existence result for a class of problems and we discuss further steps.       
\end{abstract}

\

\noindent {\bf Keywords:} 
information, data,  errors, mathematical models, manifolds, processes, dimension

\section{Introduction}
There is a vast amount of literature on the computational and mathematical methods of learning (e.g. the monograph \cite{b24}), the interaction between a computer and a human, and theoretical aspects 
of information engineering related to signal processing or control theory. Geometric methods, especially those based on Riemannian geometry, have been extensively used in the study of networks, information transmission, and data analysis. We note in particular the fundamental works of Amari, Belkin and Niyogi, and a recent series of papers by Fefferman and collaborators including \cite{feff}. More discussion and references can be found in  \cite{b3}, \cite{b4}. 

In \cite{b31}, it is stated that the three stages of development of  human intelligence are based on 

\noindent Stage 1: interaction between a person and the physical environment, (e.g. concluding that the fire is hot);  

\noindent  Stage 2: primitive interaction between a person and another human (e.g. repeating another person's actions such as drinking water); 

\noindent  Stage 3: a higher level interaction among humans based on  {\it language}, i.e. exchange of information that involves encoding and decoding processes.

The author of \cite{b31} draws parallels between these stages and machine learning. In particular, he compares Stage 1 to unsupervised learning. 

In Stage 3, the encoded information exchanged among people would be primarily about the physical world, themselves (as part of this world), and related processes. By now, humans created a huge repository of data, in various formats. This information/data is typically supported by an object in a physical world (e.g. a printed book, a painting, an mp3 music file physically stored on a computer drive). Communication among people now involves  new information that addresses (encodes) only information created by people. It is information about information (about abstract information, rather than about physical world). As examples, that includes techniques of solving equations, certain algorithms, or machine learning methods.  

In this paper, we state that the physical world is a submanifold of a manifold $W$, and the information space is a submanifold of a manifold $E$. We discuss the interactions (mappings) between the two. 
This approach resonates with a number of other perspectives such as 
\begin{itemize}
\item  G\"ardenfors' conceptual spaces, with his emphasis on geometric structures \cite{b10}, see also a more practical development in this general direction e.g. in \cite{b25}, 
\item digital twins (a general perspective in \cite{b21}, and an example of application of digital twins to a practical problem in medicine in \cite{b16}), 
\item  world/foundation models \cite{b6}, \cite{b14},   
\item  generally, mathematical modeling (see \cite{b13} for a discussion of how mathematical modeling differs from digital twins).
 \end{itemize}

Predicting, analyzing and correcting errors (or malfunctions) is an essential part of data analysis and  engineering.  There is an extensive theory or error correction and fault tolerant computation. This massive variety of methods originated from the necessity to deal with the noise and the accumulation of errors that occur as the data packets are transmitted through various types of channels. The word "error", in the most general sense, can be understood as an issue to be fixed or a problem to be addressed. In the abstract sense, it can be parameters or data that need to be modified. 
Broadly, one can address the question of faulty signal propagation in a network, regardless of the type of the network. It could be a mathematical model for a network of neurons or nerves in a living organism, or for a societal network. 
In the daily functioning of human society, there are quantifiable errors in the information flow that may cause operational issues. For example, an office assistant may accidentally communicate a wrong meeting room number (a simple email typo), or a post office worker may place an envelope in a wrong mail box (as a result of being momentarily distracted). 

There is extensive literature on error correction, including the foundational Shannon's paper \cite{b22} and von Neumann's paper \cite{b26}. While it is a difficult task to give a proper literature overview, we will mention a few other works. 
There has been a lof of effort to construct error correcting codes \cite{b9}, \cite{b17}.   
Quantum error correction, with references, is discussed in \cite{b11} and Chapter 10 of \cite{b19}.

We also include a few examples of works that address specific real life systems. The paper 
\cite{b20} discusses the contributing factors to errors in machining and presents theoretical and quantitative analysis of the model that leads to improved accuracy.  
The paper 
\cite{b23} reviews the issues that occur in large networks, such as congestion, delays or packet loss, and possible approaches to gaining control over these problems.   
The paper \cite{b7} addresses the problem of low power in radio frequency energy harvesting systems.
Mathematical representations or models of real life systems often serve as useful approximations. For instance, in \cite{b28}, the model for an electrical vehicle system allows to compare and evaluate the system responses.     

\section{Main theorem and proof}
In this section, we describe our mathematical model. 

\begin{figure}[tb]
\centering
\includegraphics[width=2.5in]{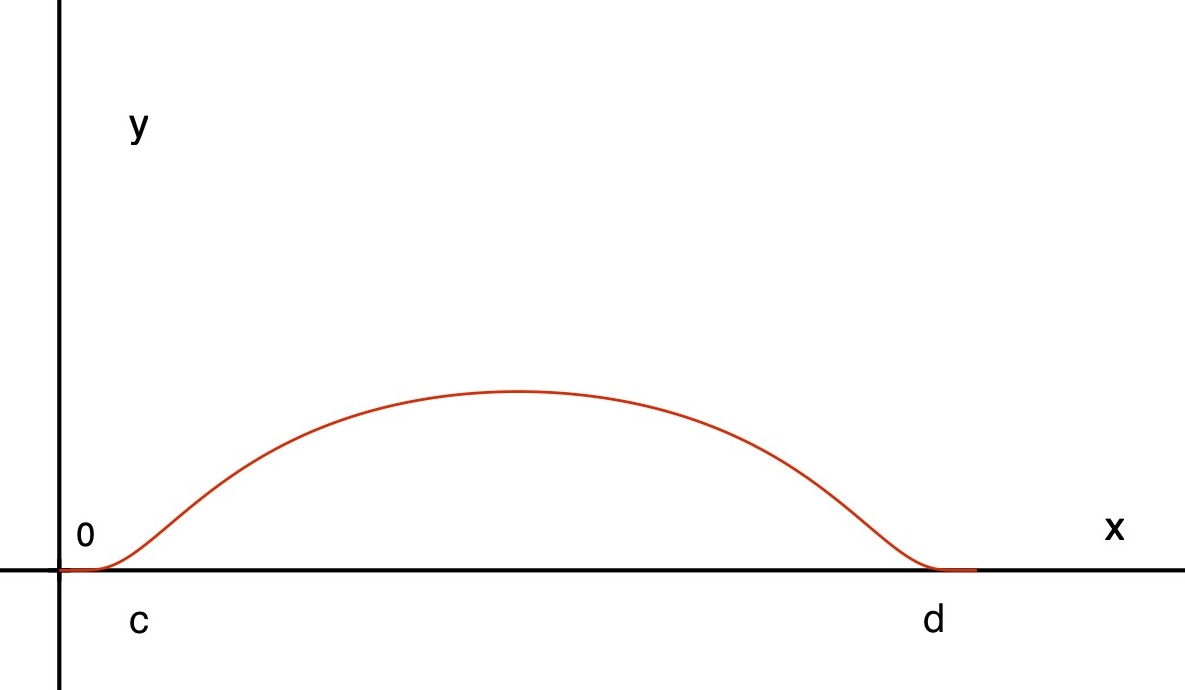}
\caption{} \label{figbump}
\end{figure}

Let H be an infinite-dimensional Hilbert space (real or complex). If H is separable, then it is unique up to an isomorphism. Equip H with the structure of a Hilbert manifold (\cite{b8}, \cite{b12}). For our analysis, let the physical world be a submanifold of a manifold W=H, and let the information space be a submanifold  of the manifold E=H. We will model a person (an individual human being) by a manifold M, together with a probability measure, p, on M. In a one-parameter smooth family (M(t),p(t))  of such pairs (we can  assume this to be a Bayesian process), where t is time, the choice of the probability measure p=p(t) is a "randomizer", which, informally speaking, reflects the unpredictability of the human choices. We assume the pair (M(t),p(t)) at every time t is represented as an  embedded submanifold of W (i.e. the choice of p is incorporated in the embedding). As a simple example, in Fig. \ref{figbump}, a manifold (c,d) (the open interval $c<x<d$ in the x-axis), with a probability distribution p whose graph is shown in the picture, is embedded into the 2-dimensional plane as the red curve, the graph of p (rather than a line segment). This is for one fixed moment of time, say t=0.  

Generally, following the spirit of \cite{b3}, \cite{b4}, whenever possible, assume that everything relevant is already contained in the manifold (or parameter space) and it is unnecessary to deal with a choice of extra structure (e.g. a metric). For example, for the manifold $J=(c,d)$ (the line segment in Fig. \ref{figbump}) equipped with a choice of a metric $\rho$, a function on $J\times J$ that assigns to each pair $(x_1,x_2)$ the value of the distance between them, 
$\rho(x_1,x_2)$, we can replace the pair $(J,\rho)$ by the set of points in the xyz-space 
$$
X=\{ (x,y,z)| \ c< x< d; c< y< d; z=\rho(x,y)\}.
$$     

\

\noindent Consider the semigroup G formed by mappings 

$U \to V$

$V \to  C \to U$ 

$V \to C_1 \to ... \to C_k  \to U$, 

\noindent where $U$, $V$ are (any) submanifolds of W, and $C$, $C_1$,..., $C_k$, for a positive integer k, are (arbitrary) submanifolds of E. These maps include, in particular, the maps that represent direct human-to-human interactions, or the human-to-human encoded communication of any form, or a person putting data into E or retrieving information from E.  
This is consistent with regarding an information transfer as an embedding of manifolds (see e.g. the discussion in 
\cite{b3}).

Everything in the information space is a result of processes in someone's neuronal networks.  Joining the information space to the physical world, in some sense,  we are looking at the collective insides of our brains.     A car, a bicycle, sports (e.g. basketball), - all this originated in someone's mind, mapped  to E, as a  set of images and other data, and then realized in W, via the semigroup G. 

\noindent {\bf Example}. In this example, we will provide a concrete example of an element of G and how it contributes to error correction. 
More examples and discussion are in Section \ref{sec:discu} below. 

Suppose Alice has to upload her Math 101 homework to the course site. She noted that assignment 1  is due October 1 (let's say, this date is $d$, on the $d$-axis in a copy of $\Bbb{R}$ embedded in the information space), Alice retrieved this information from $E$ (by checking the homework sheet) and stored the value of $d$ (in her memory cells, or in her calendar). However, being in a rush, Alice missed that the deadline is 3:00 pm on October 1. 
On October 1, Alice attempted to upload her file homework1.pdf to the course site.   
This is a map $f_1:U\to V$ of submanifolds of $W$, that induces a map $f_2:V\to C_1\subset E$, the deadline check 
map $f_3:C_1\to C_2$ in $E$ that indicates that the submission arrived after the 3:00 pm deadline, and the resulting map 
$f_4:C_2\to W$ that produces a message to Alice in $W$ that  the submission is closed and her paper can not be uploaded. The element of $G$ is $f:U\to W$ defined by the composition  $f=f_4\circ f_3\circ f_2\circ f_1$. 
A way to fix the issue, for the next assignment submission, is to increase the number of parameters, and for Alice to retrieve and store two parameters instead of one: $d$, the date, and $t$, the time (in this case, 15:00 hours).

\

The term "information space" that we chose for our submanifold of E is not the same as "information manifold" used in information geometry (see e.g. \cite{b1}). In information geometry, an information manifold (or a statistical manifold) is a set of probability measures on a measure space, equipped with additional structure (e.g. Fisher information metric).  It is possible, under appropriate assumptions, to make it a Hilbert manifold \cite{b18}. But this is not necessarily the manifold H in our context. Our H is an abstract infinite-dimensional Hilbert space with an infinite-dimensional manifold structure.     

Consider the question of fixing a malfunction (or correcting an error) in W. This is a broad question that can include actions such as replacing a failing battery or giving a person a medication.

\noindent {\bf Theorem 1.}
Let $W$, $E$ be as defined above. Let $a<b$ be real numbers and let  $I=[a,b]$. Let $n>1$ be the dimension of $S$, the submanifold of $W$ that models the physical world. Let $A$ be a submanifold of $S$. Assume the manifolds $S$, $A$ are compact,  with or without boundary.  Let $B$ be a finite-dimensional submanifold of $W$, diffeomorphic to $A$, 
such that $A\cap B=\emptyset$.  
Then there exist a submanifold $X$ of $W$ of dimension $n\le \dim X\le 2n+1$  and a smooth family of submanifolds $Y(t)$, $t\in I$, of $X$ such that $Y(a)=A$ and $Y(b)=B$. 

{\bf Proof}. We have: $\dim A\le \dim S$, $\dim B=\dim A$.  Since $S$ and $B$ are compact finite-dimensional submanifolds of $W$ and $\max \{ \dim S, \dim B\}=n$, it follows that there exists a compact n-dimensional submanifold $K$ of $W$ that contains both $S$ and $B$. By the Whitney embedding theorem (\cite{b29}, Theorem 1), there is a smooth embedding $f$  of $K$ into the $(2n+1)$-dimensional Euclidean space. Moreover,  $f(A)$ and $f(B)$ are embedded submanifolds of ${\mathbb{R}}^{2n+1}$. Let's identify this ${\mathbb{R}}^{2n+1}$ with a finite-dimensional linear subspace of $E$.  
We have: $f(K)$ is compact in ${\mathbb{R}}^{2n+1}$ (image of a compact set under a continuous map), 
$f(A)\subset f(K)$, $f(B)\subset f(K)$ and  $f(A)$ is diffeomorphic to $f(B)$. 
By the isotopy version of the Whitney embedding theorem (\cite{b30} if $\dim A=n$,  Theorem 6/deformation theorem in \cite{b29} if $\dim A<n$) there is  a smooth family of submanifolds $\Lambda (t)\subset {\mathbb{R}}^{2n+1}\subset E$, \  
$a\le t\le b$,  such that $\Lambda(a)=f(A)$ and $\Lambda(b)=f(B)$. Let $V$ be an open $(2n+1)$-dimensional ball that contains $f(K)$ and the compact manifold 
$$
\{ (t,x)| \ a\le t\le b; \ x\in \Lambda(t)\}.
$$
Next, we claim that there is a smooth map $F:V\to W$ such that $F\Bigr |_{f(K)}=f^{-1}$. This follows from the proof of Lemma 2.2/extension lemma in \cite{b15}.  Lemma 2.2 can not be used as stated in \cite{b15}, because $W$ is infinite-dimensional, and Lemma 2.2. is for maps with codomain ${\mathbb{R}}^m$. Note that we can not assume that there exists a finite-dimensional linear subspace of $W$ that contains $S$ or $K$ (this may not be true). However, the proof of lemma 2.2 \cite{b15} can be adjusted, to the case of codomain $W$, to conclude that the map $f^{-1}:f(K)\to K$, regarded as a function $f(K)\to W$,  extends to $F: V\to W$. We observe: $F(f(A))=A$, $F(f(B))=B$. 
We set $X=F(V)$ and $Y(t)=F(\Lambda(t))$ for each $t\in I$. The desired conclusion follows.   
$\Box$

\begin{figure}[htbp]
\centerline{\includegraphics[width=2.5in]{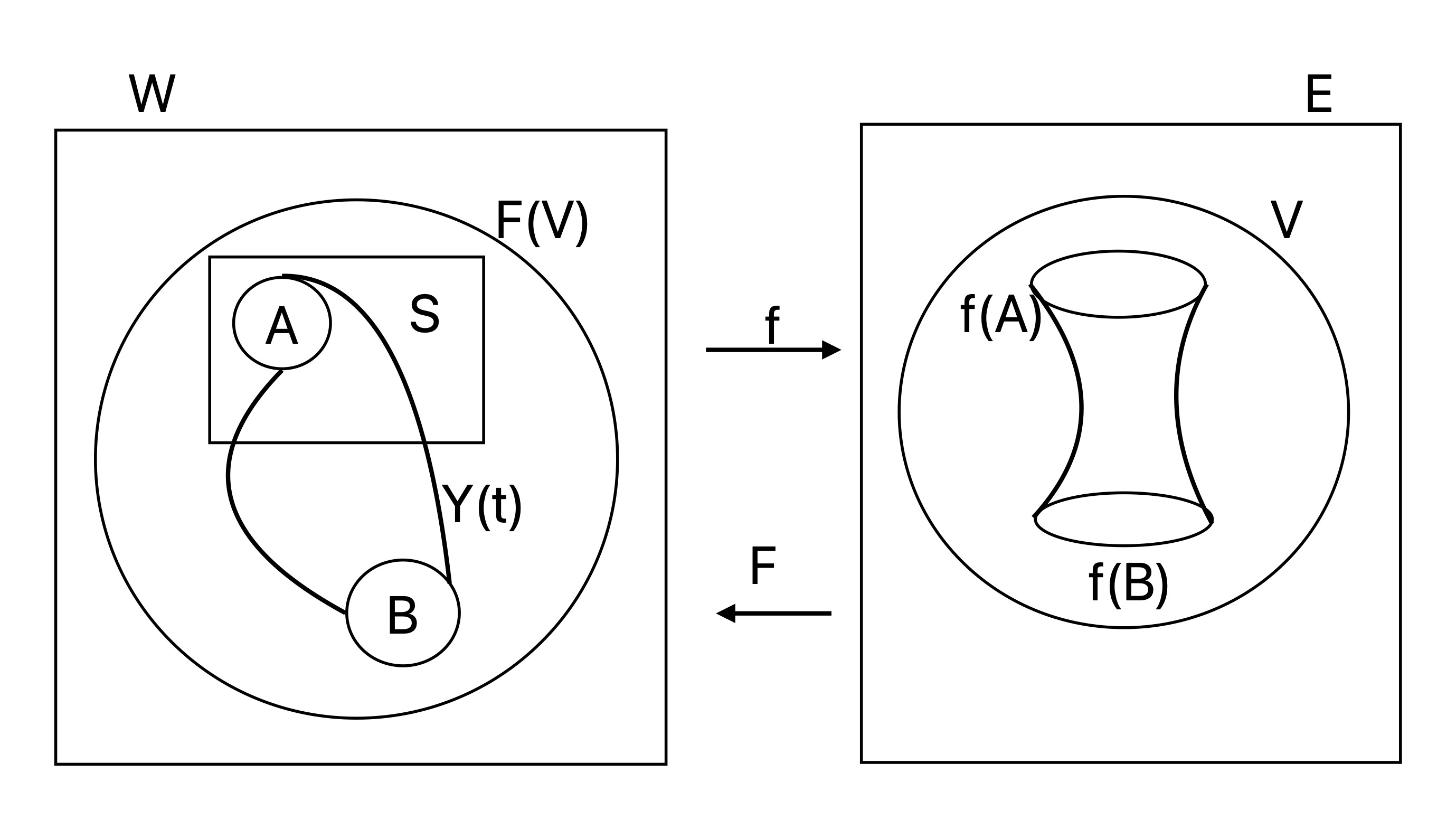}}
\caption{}
\label{figmapw}
\end{figure}

The sketch in Fig. \ref{figmapw} illustrates the idea of the proof. 

\section{Discussion}
\label{sec:discu}
In this section, the example in subsection \ref{sec:deconv} is directly relevant to  Theorem 1. 
Subsections \ref{sec:bflip}, \ref{sec:tenb}, \ref{sec:penb} are meant to illustrate at an intuitive level how solving a 
problem is tied with increasing the number of parameters in the mathematical model that approximates this system. 
\subsection{}
\label{sec:deconv}
Deconvolution is a mathematical technique that sharpens blurry images. Typically, there is a signal  
 recorded with an optical instrument (e.g. a microscope).  This recorded signal is  a function $\psi$ (the blurry image), and  $\psi=\alpha\ast \beta$, where 
 $\beta$ is a point spread function. The objective is to recover the true signal $\alpha$ (the image with a good resolution and contrast). There are various software packages commonly used for practical purposes, e.g. Fiji/ImageJ. 
 
 Applied to this situation, Theorem 1 can be interpreted as follows. Let A be a blurry photo image (represented 
 by the signal $\psi$). The theorem asserts that there must exist a process (which possibly involves doubling the number of variables, plus one more variable) which results in a more clear photo, B. In our setting, B is represented by $\alpha$.  One possible way to get from A to B is to by implementing deconvolution. 
 Using the Fourier transforms amounts to doubling the number of variables. One extra variable is time. 
 This is consistent with the assertion of the theorem.       

\subsection{} 
\label{sec:bflip}
In a very basic information processing example, a bit flip error in a specific transmission of a string of 0s and 1s can be diagnosed by sending multiple copies of the same string through the same channel.  In the quantum information set-up, this trick no longer works, due to the no-cloning theorem, but there are other ways  to detect the bit-flip errors via increasing the number of qubits. Without going into the general theory of error correction, let us simply note that in both these examples, the intent to correct the error led to increasing the size of the space of states. 
\subsection{}
\label{sec:tenb}
\begin{figure}[htbp]
\centerline{\includegraphics[width=1.2in]{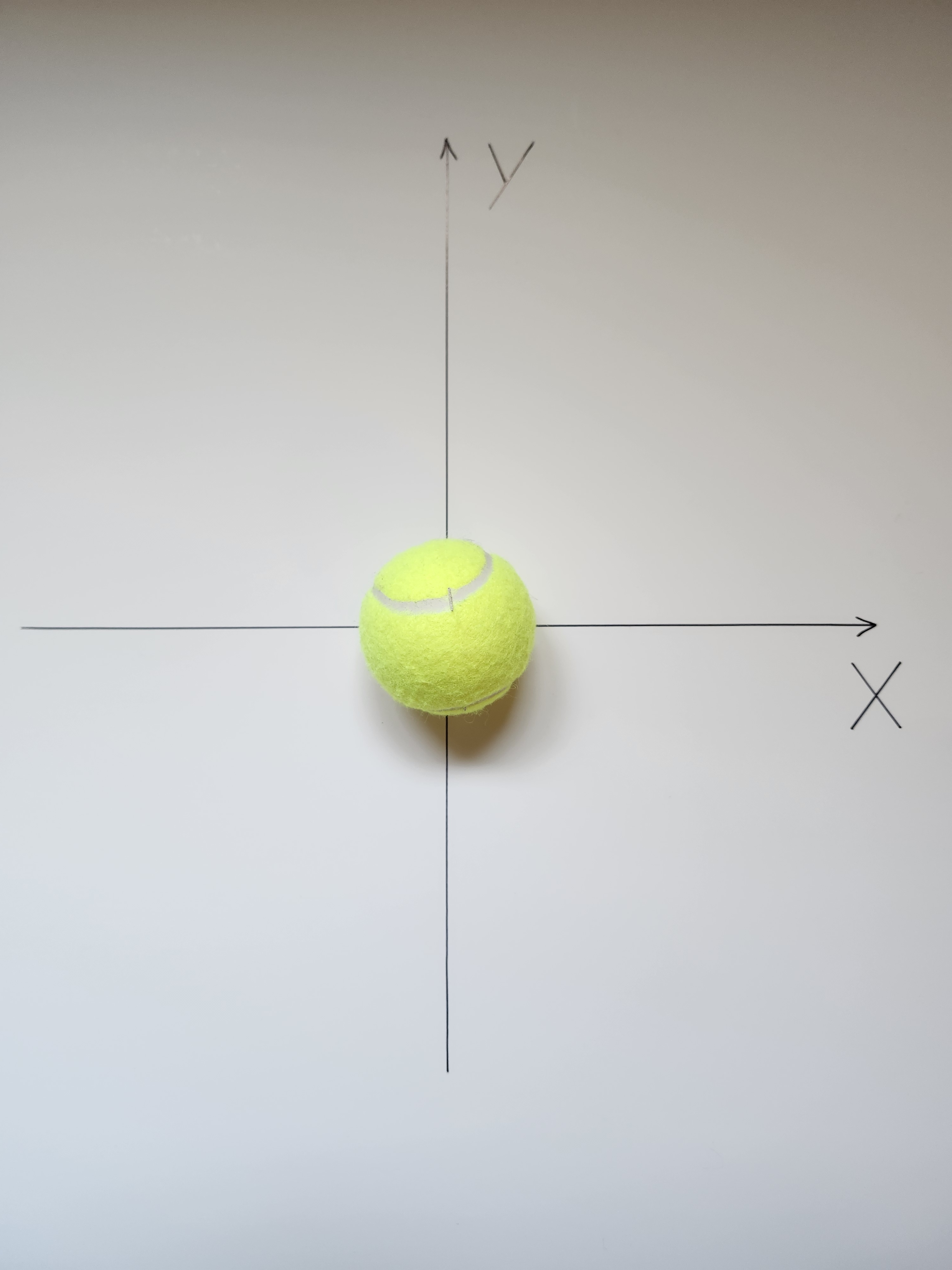}}
\caption{}
\label{figb1}
\end{figure}

\begin{figure}[htbp]
\centerline{\includegraphics[width=1.2in]{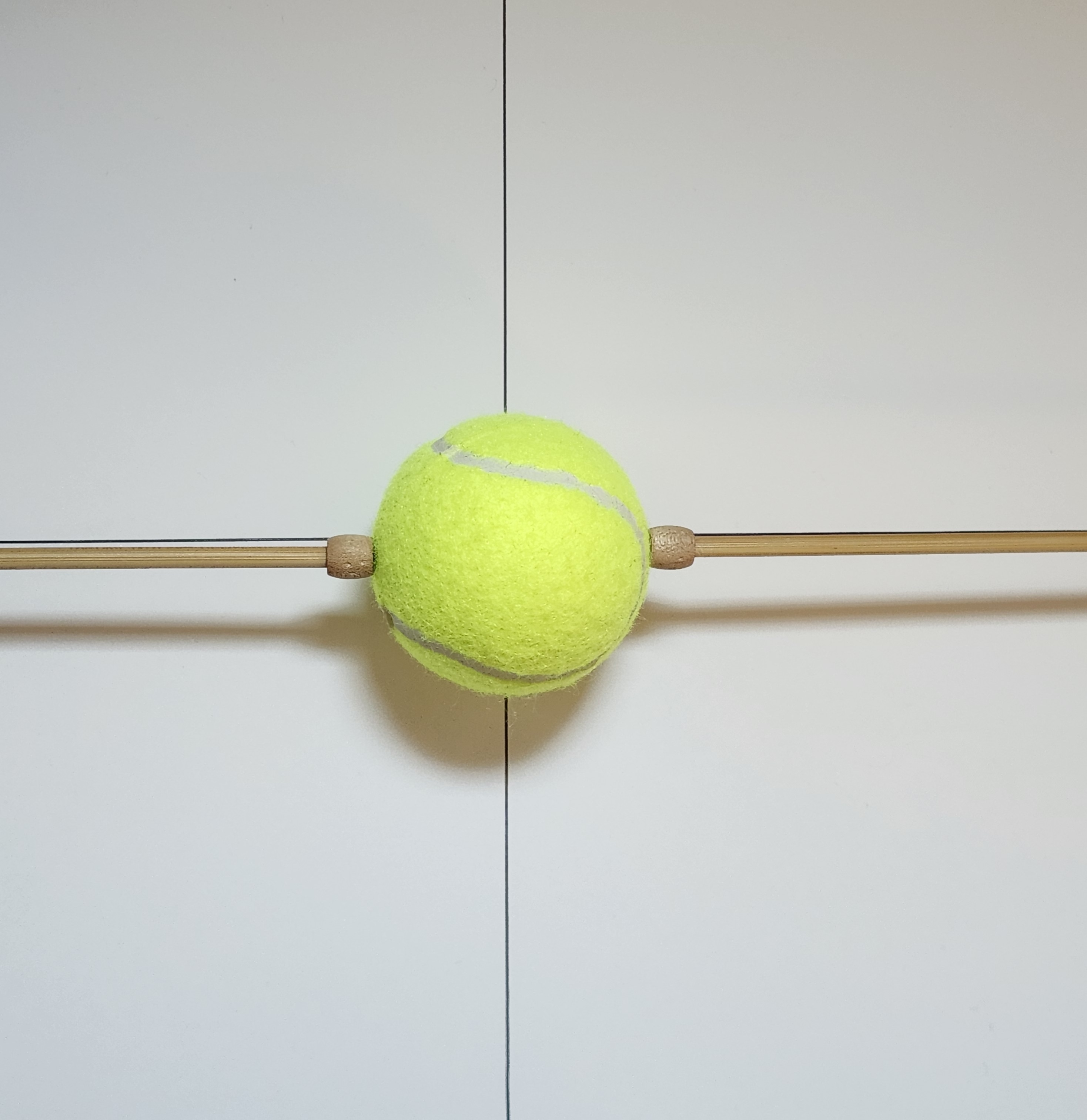}}
\caption{}
\label{figb2}
\end{figure}

\begin{figure}[htbp]
\centerline{\includegraphics[width=1.2in]{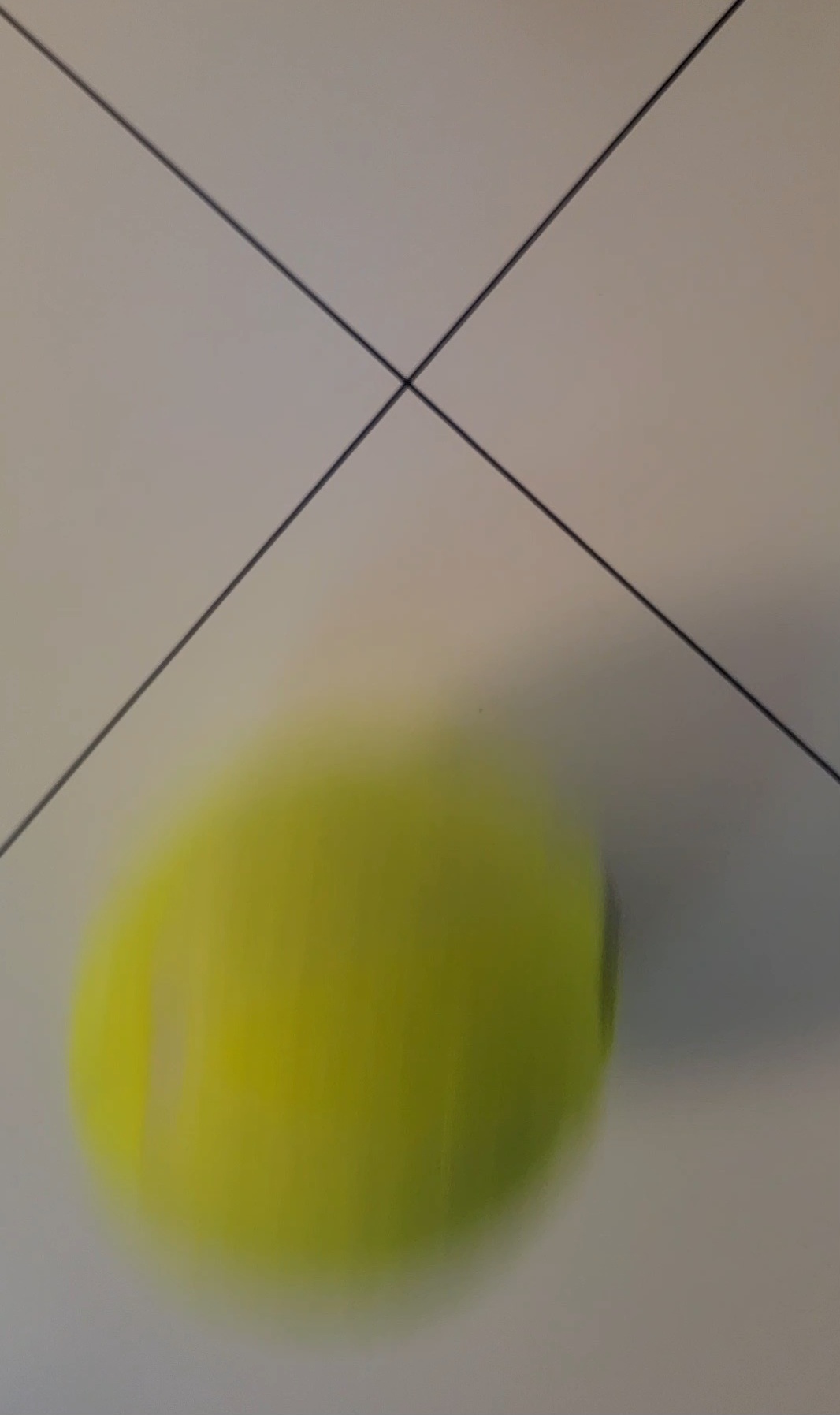}}
\caption{}
\label{figb3}
\end{figure}

In Fig. \ref{figb2}, there is a tennis ball held by two rods (see also Fig. \ref{figb1}). The xy-plane is vertical. 
The photo on Fig. \ref{figb3} is taken after the ball is released. Suppose we predicted that upon being  released, the ball would move in the negative direction of the y-axis. If we interpret it as a prediction error, then one way to  correct it would be to add one more parameter to Fig. \ref{figb1}: the direction of the gravitational force, and this would allow us to make the correct prediction for the direction of the motion of the ball. 
Another way to add parameters to the situation would be to place a mechanical arm that would hold the tennis ball and, at time t=0 would move it in the prescribed direction (in this case, in the direction of the  negative y-axis).  

\subsection{}
\label{sec:penb}
\begin{figure}[htbp]
\centerline{\includegraphics[width=1.2in]{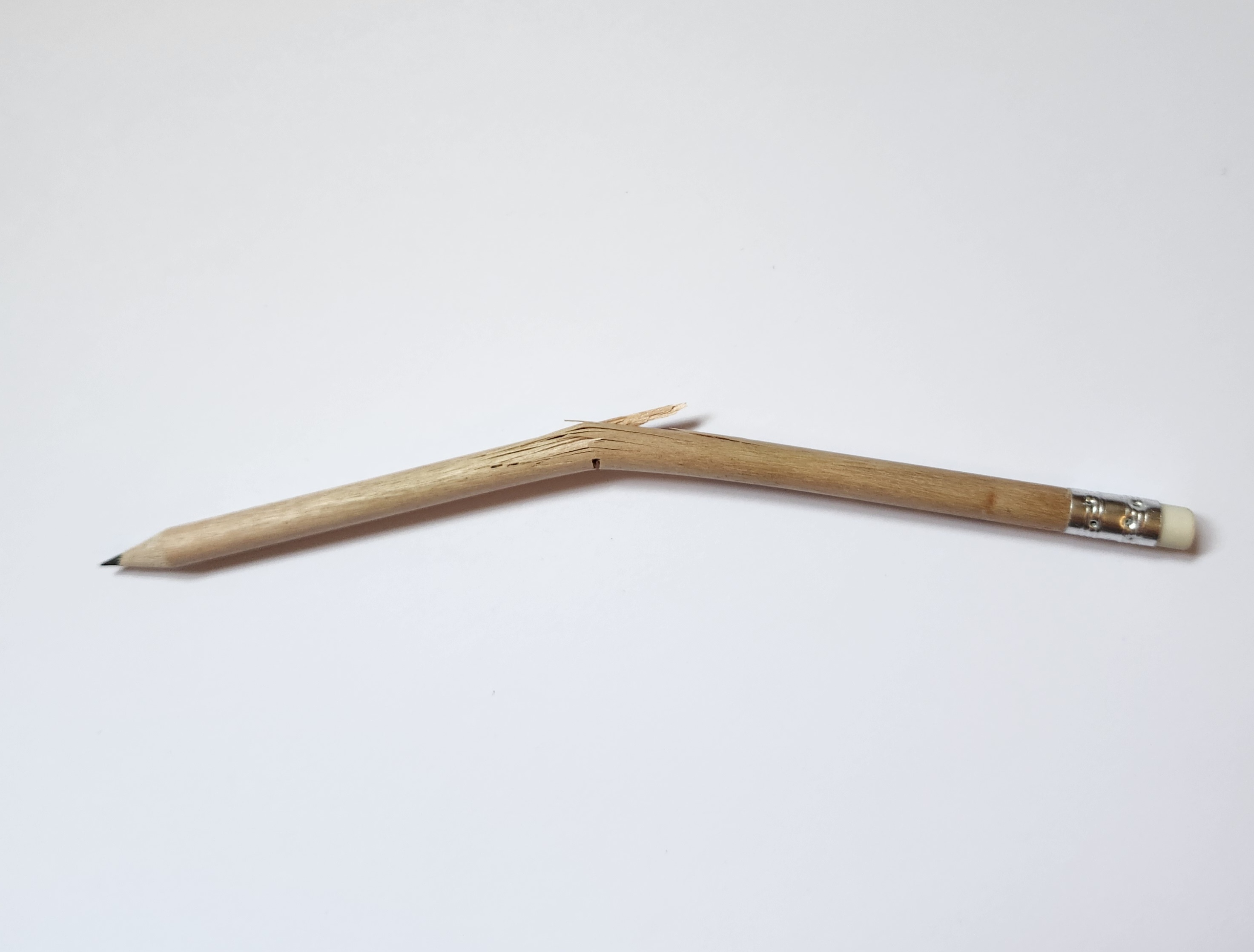}}
\caption{}
\label{figp1}
\end{figure}

\begin{figure}[htbp]
\centerline{\includegraphics[width=1.2in]{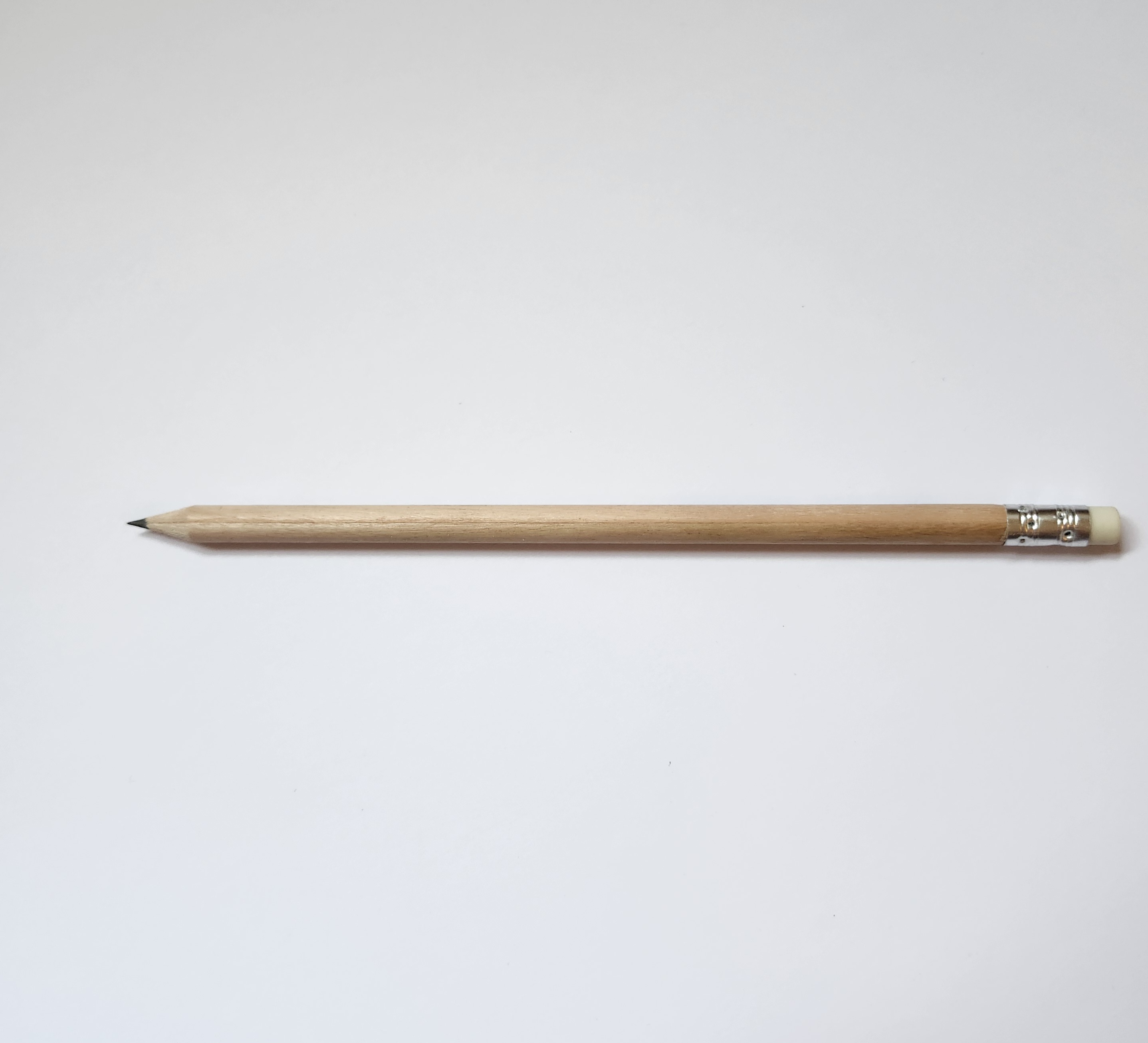}}
\caption{}
\label{figp2}
\end{figure}

In Fig. \ref{figp1}, there is a photo of a broken pencil. We can edit the image to "repair" the pencil: Fig. \ref{figp2}.  This amounts to rearranging the pixels in a 2-dimensional submanifold of E. Mirroring this process in W means putting the two pieces together. However, the actual pencil in W is still broken. The bonds at the atomic scale have not been repaired.  
The problem needs to be formulated with more parameters, to be solved in W.   

\section{Conclusions}

\subsection{}
The information space is distinct  from the physical world. We model both by submanifolds of  infinite-dimensional manifolds W and E, respectively. The information space contains the information created by processes in neural networks of human brains. Many processes can be modelled by mappings between subsets of W and subsets of E. For example, the idea of using digital twins to solve problems in the physical world fits perfectly in this setup. 
This direction of thinking is also consistent with  world models in artifical intelligence.   

Theorem 1 addresses the following situation. We consider submanifolds A and B of W, where A is in S, which we take to be the model of the physical world, or a relevant subset of the environment. The manifold S may contain, for instance, submanifolds that represent humans or other processes. 
The submanifold A  represents a set of parameters that contains an "error" (a localized data set that needs to be changed) and B the target set of parameters where this error is not present. 
As an informal comment, in practice, fixing one local error may change the values of other parameters elsewhere, but we assume that B is an acceptable outcome.   
We consider the Bayesian process in S, over the time interval I=[a,b], starting with A at t=a. Because it is governed by the laws included in the definition of S, the process will finish at a submanifold of S at time t=b. 
The theorem does not require that B is a subset of  S. The assumption is that B is a submanifold of W. If A is allowed to evolve on its own, within the limitations set by the model and the present conditions, then the outcome B may not be reachable. The theorem asserts that it is possible to find a smooth path from A to B, and that  
is comes with increasing the number of parameters used to describe the physical world, S, from n parameters  to 
(up to) 2n+1 parameters. Informally speaking, this maybe be regarded  increasing the magnification/resolution.  In other words, the theorem says that the error (or a malfunction) can be fixed by increasing the dimension of the model, but the theorem does not suggest how to choose these additional parameters (up to n+1 of those).  The proof indicates that the extra parameters appear from 
mappings between subsets of W and subsets of E, i. e. elements of the semigroup G, discussed earlier. 
However, the proof does not specify which specific elements of G give these extra parameters. 
In this sense, the proof is not constructive and  the theorem is an existence statement.

In Section \ref{sec:discu}, 
we present several examples that echo the thinking behind this theorem. The discussion  
of  deconvolution in image processing demonstrates the statement of the theorem most precisely among these.   

The theorem is meant to be a general statement/principle, applicable to mathematical models of image processing or signal processing in biomedical systems, or other systems or networks.  

\subsection{Future work and limitations.} 

This paper is an observation that ramified from \cite{b2}, \cite{b3}, \cite{b4}. The general line of thought (that in some sense has roots in \cite{b26}, \cite{b27}) 
 is to develop the mathematical methods allowing to deal with the issues that occur with electrical signal transmission in human body. The question addressed in this note is about solvability of the problem, and the answer is an existence statement for a solution (Theorem 1). This answer is not constructive, as it is only an assertion that a solution exists, and this can be regarded as a limitation. Another limitation is the setup of the mathematical model (in particular, the requirement that $B$ is diffeomorphic to $A$, even though we do not require that $B$ is a subset of $S$, which allows for some freedom, as it means that $B$ describes a configuration that is currently not realized in the physical world).   

It is known that a number of issues with the human body (for example, certain instances of chronic pain) result from 
a malfunction in neurons, although there is no complete knowledge of these mechanisms or of the ways to repair it. A suitable direction for future work is towards a mathematical model that addresses the geometry of this situation and the appropriate choice of the parameter space. To handle the uncertainty, one possible approach is to use fuzzy manifolds, or a suitable generalization of this concept. This should work well with hidden Markov processes or quantum hidden Markov processes.  
Quantumness is a separate issue that needs to be addressed, as the signal propagation in neural networks 
in human body seems to be a "partially classical and partially quantum" process.  
Replacing a probability measure $\rho$ on a manifold $M$ with a smooth membership function $\mu:M\to [0,1]$ would change our interpretation of $M$. In a simple example  
when $M$ is the open unit disk in the $xy$-plane and $\rho=\frac{1}{\pi}dA$, where $dA$ is the Lebesgue measure,   
there are many constant membership functions: $\mu_c:M\to [0,1]$, $\mu_c(x,y)\equiv c$, where $c$ is a constant between $0$ and $1$. The function $\mu_1$ corresponds to the case when for every point $(x,y)$ 
such that $x^2+y^2< 1$ we have 
$(x,y)\in M$ in the fuzzy set $(M,\mu_1)$.   The function $\mu_0$ corresponds to the case when for every point $(x,y)$ 
such that $x^2+y^2< 1$, 
$(x,y)\notin M$ in the fuzzy set $(M,\mu_0)$.   
Working with fuzzy manifolds is difficult, as it very quickly becomes clear already at the level of 
fuzzy topological spaces \cite{b5}.  
But, intuitively, this can be a right mathematical concept to support the study of quantitative processes that happen for an unknown reason (although the reason can possibly be identified in the correctly chosen parameter space).    

\section*{Acknowledgment}

T.B. thanks the anonymous reviewers for their valuable comments and suggestions.

\end{document}